\documentclass[pdflatex,sn-mathphys-num]{sn-jnl}

\usepackage{graphicx}%
\usepackage{multirow}%
\usepackage{amsmath,amssymb,amsfonts}%
\usepackage{amsthm}%
\usepackage{mathrsfs}%
\usepackage[title]{appendix}%
\usepackage{xcolor}%
\usepackage{textcomp}%
\usepackage{manyfoot}%
\usepackage{booktabs}%
\usepackage{algorithm}%
\usepackage{algorithmicx}%
\usepackage{algpseudocode}%
\usepackage{listings}%

\theoremstyle{thmstyleone}%
\theoremstyle{thmstyletwo}%

\theoremstyle{thmstylethree}%

\begin{document}

\title{SIGMA: speaker-isolated, symptom-structured multimodal estimation of depression severity from clinical interviews}


\author[1,2]{\fnm{Zhuoyuan} \sur{Tang}}

\author*[1]{\fnm{Tao} \sur{Wang}}
\email{tao.wang@kcl.ac.uk}

\author[3]{\fnm{Kai} \sur{Yang}}

\author[4]{\fnm{Li} \sur{Yuan}}

\author[1]{\fnm{Yuezhou} \sur{Zhang}}

\author[1,5]{\fnm{Rongrong} \sur{Zhong}}

\author[4]{\fnm{Yi} \sur{Cai}}

\author[1]{\fnm{Saskia C.} \sur{Sanderson}}

\author[6]{\fnm{Jaime} \sur{Delgadillo}}

\author[1,7,8]{\fnm{Richard} \sur{Dobson}}

\author[1]{\fnm{Angus} \sur{Roberts}}

\affil[1]{\orgdiv{Department of Biostatistics and Health Informatics}, \orgname{King's College London}, \orgaddress{\city{London}, \country{UK}}}

\affil[2]{\orgdiv{Department of Intelligence and Computing, School of Computer Software}, \orgname{Tianjin University}, \orgaddress{\city{Tianjin}, \country{China}}}

\affil[3]{\orgdiv{College of Economics}, \orgname{Shenzhen University}, \orgaddress{\city{Shenzhen}, \country{China}}}

\affil[4]{\orgdiv{School of Software Engineering}, \orgname{South China University of Technology}, \orgaddress{\city{Guangzhou}, \country{China}}}

\affil[5]{\orgdiv{Shanghai Mental Health Center, School of Medicine}, \orgname{Shanghai Jiao Tong University}, \orgaddress{\city{Shanghai}, \country{China}}}

\affil[6]{\orgdiv{Department of Psychology}, \orgname{King's College London}, \orgaddress{\city{London}, \country{UK}}}

\affil[7]{Institute of Health Informatics, University College London, Euston Road, London NW1 2DA, United Kingdom;}

\affil[8]{Health Data Research UK London, University College London, Euston Road, London NW1 2DA, United Kingdom;}


\abstract{Clinical interviews are central to depression assessment but time-consuming and rater-dependent. Automated severity estimation offers scalable screening, yet existing methods often retain interviewer speech (confounding participant cues), fuse modalities late (losing fine-grained cross-modal dynamics), or score symptoms independently (ignoring inter-symptom correlations). We present SIGMA, a speaker-aware, symptom-structured framework for depression-severity estimation from semi-clinical interviews. SIGMA uses speaker diarization to restrict input to participant speech, cross-attention so that text, audio and video condition one another during encoding, and a graded-response head that predicts the eight questionnaire items jointly through a shared latent severity trait with item-specific ordered thresholds. On the Extended Distress Analysis Interview Corpus, SIGMA achieves state-of-the-art performance (CCC 0.762, RMSE 4.13, MAE 3.30 over five seeds). Ablations confirm that isolating participant speech and joint, structured symptom prediction each substantially improve concordance and estimation accuracy.
}

\keywords{depression severity estimation, multimodal learning, speaker diarization, item response theory, multi-task learning}



\maketitle

\section*{Introduction}
\label{sec:introduction}
\addcontentsline{toc}{section}{Introduction}

Depression affects an estimated 332 million people worldwide and is among the leading contributors to years lived with disability \cite{who2025depression}. It disrupts mood, cognition and daily functioning and imposes a substantial clinical, social and economic burden \cite{lepine2011increasing}, so accurate and timely assessment is essential for effective treatment and improved outcomes \cite{oluboka2025urgency}. Current assessments often rely on clinical interviews \cite{sadeghi2024harnessing} and screening instruments such as the Patient Health Questionnaire-8 (PHQ-8) \cite{kroenke2009phq}. These methods are clinically valuable but subjective \cite{american2013diagnostic}, resource-intensive \cite{atzil2024leveraging} and vulnerable to recall and rater bias \cite{sadeghi2024harnessing}, and the burden compounds in longitudinal care, where patients repeat questionnaires or attend interviews at every follow-up \cite{kazdin2008evidence}. Automated estimation of severity from the interview itself could relieve part of that burden and make screening and longitudinal monitoring more scalable and more consistent. The difficulty is that depression manifests across language, prosody, facial expression and behaviour \cite{cameron2024multimodal}, and no single channel carries the full clinical picture.

Multimodal machine learning has made it increasingly feasible to combine these channels \cite{sadeghi2024harnessing,mandal2025enhancing}. Early work relied on social-media text, using linguistic metadata and neural encoders, semi-supervised symptom tracking, and models conditioned on PHQ items \cite{orabi2018deep,yazdavar2017semi,nguyen2022improving}. Recent research also turned to semi-clinical interview corpora, which resemble real assessment more closely: the DAIC-WOZ corpus \cite{gratch2014distress} and its extension E-DAIC (Extended Distress Analysis Interview Corpus) \cite{devault2014simsensei,ringeval2019avec} pair standardised transcripts and acoustic and visual features with PHQ-8 labels, and have become the standard benchmark for the task. Work on these corpora has progressed from single modalities, using word embeddings and BERT-style encoders for text \cite{orabi2018deep,devlin2019bert,sadeghi2023exploring} and OpenSMILE and OpenFace descriptors for audio and video \cite{eyben2010opensmile,baltruvsaitis20182,xu2023multimodal}, to multimodal systems that combine them by early concatenation \cite{rodrigues2019multimodal}, cross-modal attention \cite{tsai2019multimodal} or late fusion \cite{sadeghi2024harnessing}, and more recently to item-level formulations that predict the eight PHQ-8 questions separately \cite{van2022text,mandal2025enhancing}.

Three limitations persist across previous studies. First, clinical interviews are multi-turn interactions in which participant and interviewer speech is interspersed \cite{milintsevich2023towards,mandal2025enhancing}, yet many methods process the full recording or transcript without speaker separation \cite{sadeghi2024harnessing,mandal2025enhancing}, allowing interviewer-specific cues such as questioning style and prosody to introduce task-irrelevant variation. Second, text, audio and video provide complementary cues that clinicians weigh jointly \cite{mandal2025enhancing}, yet many methods encode each modality separately and combine only the final predictions, for example by averaging modality-specific scores \cite{sadeghi2024harnessing}. Such late fusion cannot represent cross-modal interactions, such as how a hesitant tone changes the interpretation of a spoken answer, and other approaches use only a subset of the available streams \cite{xu2023multimodal2,zhang2025optimizing}. Third, the PHQ-8 is comprised of the sum of eight related-symptom items (Q1–Q8) rather than a single Likert scale, and these symptoms co-occur: reduced interest or pleasure (Q1) commonly accompanies depressed mood (Q2). Mandal et al.\ \cite{mandal2025enhancing} showed that item-wise prediction improves total-score estimation, but their framework predicts each item independently and does not model dependencies among symptoms. Moreover, each PHQ-8 item is scored on four categories that are ordered but not evenly spaced: ``not at all'', ``several days'', ``more than half the days'' and ``nearly every day''. Classifiers treat them as unordered, and ordinal losses restore the ordering only by assuming equal spacing, so neither allows the model to learn how far apart the categories are. The categories are also imbalanced: most participants answer ``not at all’’ to most items, so an objective that weights every response equally is dominated by the majority category \cite{mandal2025enhancing}.

\begin{figure}[!ht]
\centering
\includegraphics[width=1.0\textwidth]{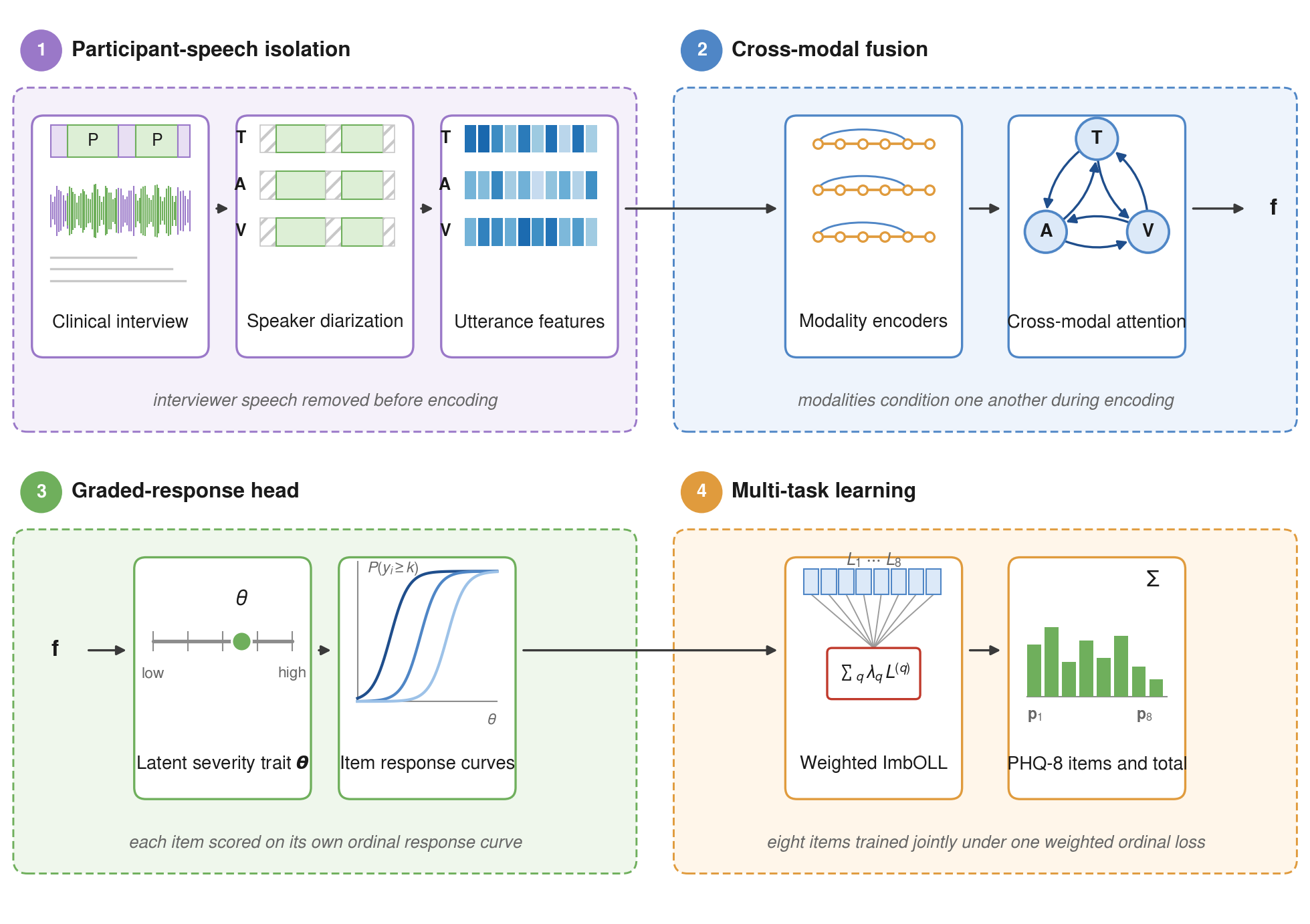}
\caption{\textbf{SIGMA, the proposed framework for multimodal depression severity estimation on E-DAIC.} The pipeline includes four components. \textbf{1, Participant-speech isolation.} Pyannote speaker diarization labels the interviewer and participant turns of each session; interviewer turns are removed from the text and audio streams, and one bge, eGeMAPS and OpenFace feature vector is computed per participant utterance. The visual stream is pooled over the same participant utterances and needs no speaker identification, since E-DAIC video shows only the participant. \textbf{2, Cross-modal fusion.} Each modality is encoded by a BiLSTM with self-attention; cross-attention lets the modalities condition one another (all six ordered pairs) before pooling into the fused representation $\mathbf{f}$. \textbf{3, Graded-response head.} A trait network maps $\mathbf{f}$ to a latent severity trait $\boldsymbol{\theta}$; each PHQ-8 item responds to the trait through its own response curve, giving eight ordinal item distributions. \textbf{4, Multi-task learning.} The eight item losses are combined into one weighted W-ImbOLL objective with learned task weights, so the items are trained jointly, and the expected item scores are summed to the total PHQ-8 score. Block-level detail is given in the Methods.}
\label{fig:overall_framework}
\end{figure}

To address these limitations, we propose SIGMA (Speaker-Isolated Graded-response Multimodal Architecture; Fig.~\ref{fig:overall_framework}), a framework with four components. First, SIGMA applies speaker diarization \cite{bredin2020pyannote,Plaquet23} to separate participant from interviewer speech before feature extraction, so that the model sees the speaker the label describes. Second, modality-specific bidirectional long short-term memory (BiLSTM) encoders and six pairwise cross-attention modules let text, audio and video condition one another during representation learning rather than only after independent processing. Third, a graded-response head \cite{samejima1969estimation} maps the fused representation to a latent severity trait and predicts each item through its own response curve, with an item-specific discrimination and ordered thresholds learned from the data, so that the ordering is enforced by the model and the spacing between categories is estimated rather than assumed. Fourth, the eight items are trained jointly under a multi-task learning objective \cite{zhang2018overview,ruder2017overview}, using a Weighted Imbalanced Ordinal Log-Loss (W-ImbOLL) that accounts for both class imbalance and ordinal error distance. We evaluate SIGMA and each of its components on E-DAIC.

Our results show SIGMA sets a new state-of-the-art on the official E-DAIC test partition, outperforming previously published results. A leave-one-out ablation under a matched evaluation protocol further identifies the sources of this performance. Participant-speech isolation and the graded-response head improve agreement and error together on both partitions, and joint training raises per-item agreement on every configuration at a fraction of the cost of eight single-item models. Cross-modal attention, by contrast, changes accuracy by less than the seed-to-seed variation. Diagnostic analyses help explain this limited contribution: permuting the audio or video stream leaves almost all predictions unchanged, and the trained cross-attention reduces to uniform pooling.
Interestingly, the graded-response head, which explicitly models item-specific response curves, reveals that affective and cognitive symptoms have greater discriminative power than somatic symptoms.

\section*{Results}
\label{sec:results}
\addcontentsline{toc}{section}{Results}


We estimated depression severity using E-DAIC \cite{devault2014simsensei,ringeval2019avec}, 275 semi-clinical interviews with US adults from the general and veteran populations, conducted by a virtual interviewer, split into the official 163 training, 56 development and 56 test sessions (Supplementary Table~S1) and distributed with audio, transcripts and pre-extracted OpenSMILE acoustic \cite{eyben2010opensmile} and OpenFace visual \cite{baltruvsaitis20182} features. The PHQ-8 records how often each of eight depressive symptoms has been present over the preceding two weeks, on four ordered response categories: 0 (not at all), 1 (several days), 2 (more than half the days) and 3 (nearly every day) \cite{kroenke2009phq}. We predict all eight item scores and sum them to the total PHQ-8 score $[0,24]$.

\begin{figure}[!ht]
\centering
\includegraphics[width=1.0\textwidth]{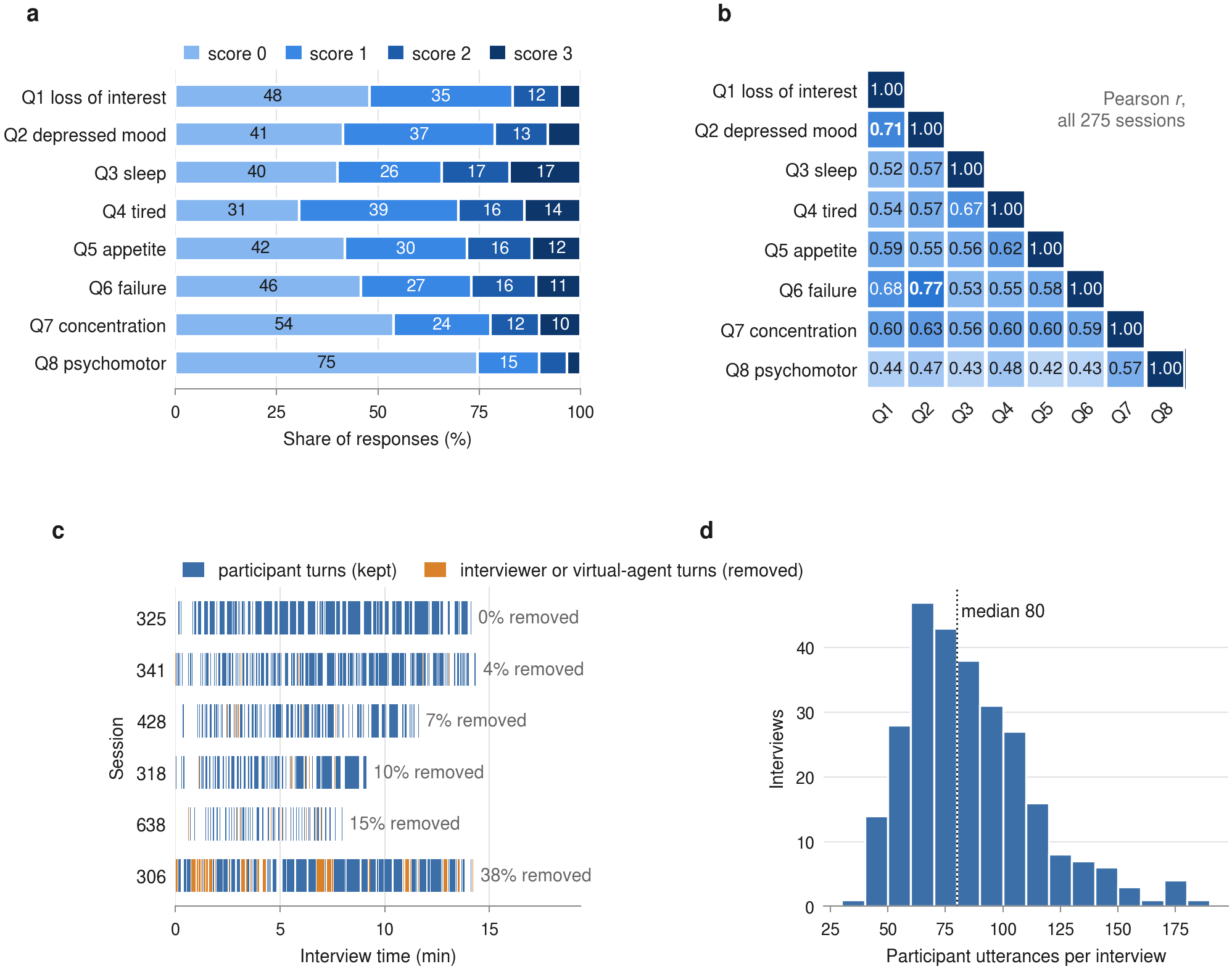}
\caption{\textbf{E-DAIC label and speech statistics.} \textbf{a}, Proportion of responses in each of the four ordinal categories for each PHQ-8 item over all 275 sessions. \textbf{b}, Pearson correlations among the eight items over all 275 sessions (lower triangle); pairs with $r \geq 0.7$ are in bold. \textbf{c}, Speaker turns in six interviews chosen at equal quantiles of the share of removed speech, each on its own time axis: blue segments are participant turns retained after diarization, orange segments are interviewer or virtual-agent turns that are removed, with the removed share given per session. \textbf{d}, Participant utterances per interview after diarization over the 275 sessions; the dotted line marks the median of 80 utterances. Participant-only sequences contain 38--190 utterances; all sequences are zero-padded to a common width of 203, the longest original transcript in the corpus, and no interview is truncated.}
\label{fig:data}
\end{figure}

Three properties of these labels shape the modelling problem. First, the four categories are ordered frequencies rather than unordered classes, so the distance between them carries meaning: answering ``nearly every day'' when the truth is ``not at all'' is a worse error than answering ``several days'', and a classifier over four nominal categories cannot represent that. Nor are the categories evenly spaced, so an ordinal loss that scores an error by the number of categories between the prediction and the true response assumes a spacing the questionnaire does not supply; the graded-response head instead places the categories on one severity scale separated by thresholds estimated from the data. Second, the categories are used unevenly in an item-specific way (Fig.~\ref{fig:data}a): score 0 is the most frequent response for every item except Q4 (feeling tired), ranging from 30.5\% of responses for Q4 to 74.5\% for Q8 (psychomotor changes), while the highest category ranges from 3.3\% to 17.5\% across items. An objective weighting all items and categories alike is therefore dominated by the majority category of the most skewed items, which motivates the item-specific inverse-frequency weighting in W-ImbOLL. Third, the items are correlated rather than independent (Fig.~\ref{fig:data}b) and internally consistent as a scale (Cronbach's $\alpha=0.911$; Supplementary Table~S1), so item-level prediction is a composite problem rather than eight separate ones, and one dimension accounts for most of what the items share. This motivates both sharing one trunk across the eight tasks and giving the graded-response head a single latent severity trait.

The proportion of each recording comprising participant speech also varies across interviews. Diarization removes a median of 8\% of transcript turns attributed to interviewers, ranging from 0 to 37\% depending on the session (Fig.~\ref{fig:data}c). The share is also unevenly distributed across the partitions, and is highest in the one used for evaluation: a median of 12.7\% of turns in the test partition against 7.4\% and 6.6\% in training and development (Supplementary Table~S1). After diarization the retained participant speech comprises 38--190 utterances per interview, with a median of 80 (Fig.~\ref{fig:data}d).

In light of these data characteristics, SIGMA incorporates four components (Fig.~\ref{fig:overall_framework}). \emph{Participant-speech isolation} applies speaker diarization to remove interviewer turns from the text and audio streams, giving one feature vector per participant utterance. \emph{Cross-modal fusion} encodes each modality with a BiLSTM and self-attention, and pairwise cross-attention modules let the modalities condition one another before pooling into the fused representation $\mathbf{f}$. The \emph{graded-response head} maps $\mathbf{f}$ to a latent severity trait and predicts each item through its own response curve. \emph{Multi-task learning} trains the eight items jointly under the W-ImbOLL objective and sums the expected item scores to the total. The Methods specify each component in full.

Following previous studies \cite{mandal2025enhancing,nerella2026clinically,zhou2025multi}, we report three metrics for the total PHQ-8 score, averaged over five random seeds: the concordance correlation coefficient (CCC), higher when agreement is stronger, and the root mean square error (RMSE) and mean absolute error (MAE) in PHQ-8 points, lower when error is smaller; the Methods give the definitions. Four questions organise the experiments, each following from the answer to the last. \textbf{RQ1} asks how accurately SIGMA estimates PHQ-8 severity relative to published E-DAIC systems, and where it fails. \textbf{RQ2} asks which inputs are required to preserve accuracy: how much of it survives when audio or video is unavailable, and what the non-verbal streams contribute when present. \textbf{RQ3} asks which parts of the framework exploit those inputs, and under what conditions each of the four components contributes. \textbf{RQ4} asks what the model has learned about the symptoms themselves, and what its fitted item parameters describe.

\begin{table}[!ht]
\centering
\caption{\textbf{Comparison with published E-DAIC results.} Methods are grouped by modelling strategy. The SIGMA row is the trimodal model, test-set mean $\pm$ s.d.\ over five random seeds on the official partition with expected-value decoding of the item scores; published rows are single reported runs. ``--'' denotes an unreported metric, arrows indicate the preferred direction, and boldface highlights the best result for each metric.}
\label{tab:compare_results}
\footnotesize
\setlength{\tabcolsep}{4.5pt}
{\begin{tabular}{lcccc}
\hline
\textbf{Model} & \textbf{Modalities} & \textbf{CCC ($\uparrow$)} & \textbf{RMSE ($\downarrow$)} & \textbf{MAE ($\downarrow$)} \\
\hline
\multicolumn{5}{l}{\textit{Group 1: conventional multimodal feature fusion}} \\
\quad Ray et al.\ (2019) \cite{ray2019multi} & Text + Audio + Video & 0.67 & 4.73 & 4.02 \\
\quad Yang et al.\ (2020) \cite{yang2020feature} & Text + Audio + Video & -- & 5.52 & 4.63 \\
\quad Sun et al.\ (2022) \cite{sun2022cubemlp} & Text + Audio + Video & 0.583 & -- & 4.37 \\
\quad He et al.\ (2024) \cite{he2024dpd} & Text + Audio + Video & 0.68 & 4.79 & -- \\
\quad Sadeghi et al.\ (2024) \cite{sadeghi2024harnessing} & Text + Video & -- & 4.53 & 3.76 \\
\quad Kumar et al.\ (2025) \cite{kumar2025multimodal} & Text + Audio & -- & 6.31 & 5.07 \\
\quad Li et al.\ (2025) \cite{li2025audio} & Audio + Video & 0.584 & 5.10 & 4.41 \\
\quad Lu et al.\ (2025) \cite{lu2025dstc} & Text + Audio + Video & -- & 5.93 & 4.85 \\
\quad Wang et al.\ (2025) \cite{wang2025multimodal} & Text + Audio + Video & -- & 5.38 & 4.16 \\
\quad Zhou et al.\ (2025) \cite{zhou2025multi} & Text + Audio & 0.60 & 5.00 & 4.05 \\
\quad Nerella et al.\ (2026) \cite{nerella2026clinically} & Text + Audio & 0.52 & 5.15 & 4.13 \\
\hline
\multicolumn{5}{l}{\textit{Group 2: richer multimodal interaction}} \\
\quad Zhang et al.\ (2020) \cite{zhang2020multimodal} & Text + Audio + Video & 0.528 & 4.47 & -- \\
\quad Sun et al.\ (2021) \cite{sun2021multi} & Text + Audio + Video & 0.331 & -- & 6.22 \\
\quad Zhao et al.\ (2022) \cite{zhao2022unaligned} & Text + Audio + Video & -- & 5.78 & 5.13 \\
\quad Saggu et al.\ (2022) \cite{saggu2022depressnet} & Text + Audio + Video & 0.45 & 5.36 & -- \\
\quad Li et al.\ (2024) \cite{li2024fpt} & Audio + Video & -- & 4.80 & 4.58 \\
\quad Shen et al.\ (2024) \cite{shen2024multi} & Text + Audio + Video & 0.563 & 4.80 & -- \\
\quad Yuan et al.\ (2024) \cite{yuan2024depression} & Text + Audio + Video & 0.676 & 4.91 & 3.98 \\
\hline
\multicolumn{5}{l}{\textit{Group 3: symptom-aware prediction}} \\
\quad Van Steijn et al.\ (2022) \cite{van2022text} & Text + Audio & 0.62 & 6.06 & -- \\
\quad Mandal et al.\ (2025) \cite{mandal2025enhancing} & Text + Audio + Video & 0.662 & 5.25 & 3.95 \\
\hline
\textbf{SIGMA} & Text + Audio + Video & \textbf{0.762\,\tiny{$\pm$\,0.021}} & \textbf{4.13\,\tiny{$\pm$\,0.17}} & \textbf{3.30\,\tiny{$\pm$\,0.06}} \\
\hline
\end{tabular}
}
\end{table}

\subsection*{RQ1: Accuracy of severity estimation and the structure of its errors}
\label{sec:results-rq1}

\textbf{SIGMA improves agreement and reduces error simultaneously.} SIGMA estimates PHQ-8 severity more accurately than any published E-DAIC system we could compare against, and on all three metrics at once (Table~\ref{tab:compare_results}). The trimodal model reaches CCC 0.762, RMSE 4.13 and MAE 3.30 over five seeds, against the best published value for each metric, each from a different system: CCC 0.68 \cite{he2024dpd}, RMSE 4.47 \cite{zhang2020multimodal} and MAE 3.76 \cite{sadeghi2024harnessing}. This is a relative improvement of 12.1\% in CCC and relative reductions of 7.6\% in RMSE and 12.2\% in MAE. Published systems trade the two against one another (Table~\ref{tab:compare_results}): the lowest reported RMSE (4.47) comes with a CCC of 0.528, and the highest reported CCC (0.68) with an RMSE of 4.79. SIGMA improves both. 
Among the three families of baselines in Table~\ref{tab:compare_results}, conventional feature-level and attention-based fusion \cite{ray2019multi,yang2020feature,sun2022cubemlp,he2024dpd,sadeghi2024harnessing,kumar2025multimodal,li2025audio,lu2025dstc,wang2025multimodal,zhou2025multi,nerella2026clinically} reaches CCC 0.52--0.68; the richer hierarchical, graph-based and higher-order mechanisms \cite{zhang2020multimodal,sun2021multi,zhao2022unaligned,saggu2022depressnet,li2024fpt,shen2024multi,yuan2024depression} reach 0.676 at best \cite{yuan2024depression}; and the symptom-level approaches \cite{van2022text,mandal2025enhancing} to which SIGMA belongs reach 0.62 and 0.662.
Notably, a number of earlier approaches introduce external dependencies as well. For example, Sadeghi et al.\ \cite{sadeghi2024harnessing} reached RMSE 4.53 and MAE 3.76 by extracting depression indicators from the transcripts with a proprietary large language model (GPT-3.5) and combining them with facial-behaviour features. SIGMA improves on both figures using openly available components run locally.

\begin{figure}[!ht]
\centering
\includegraphics[width=1.0\textwidth]{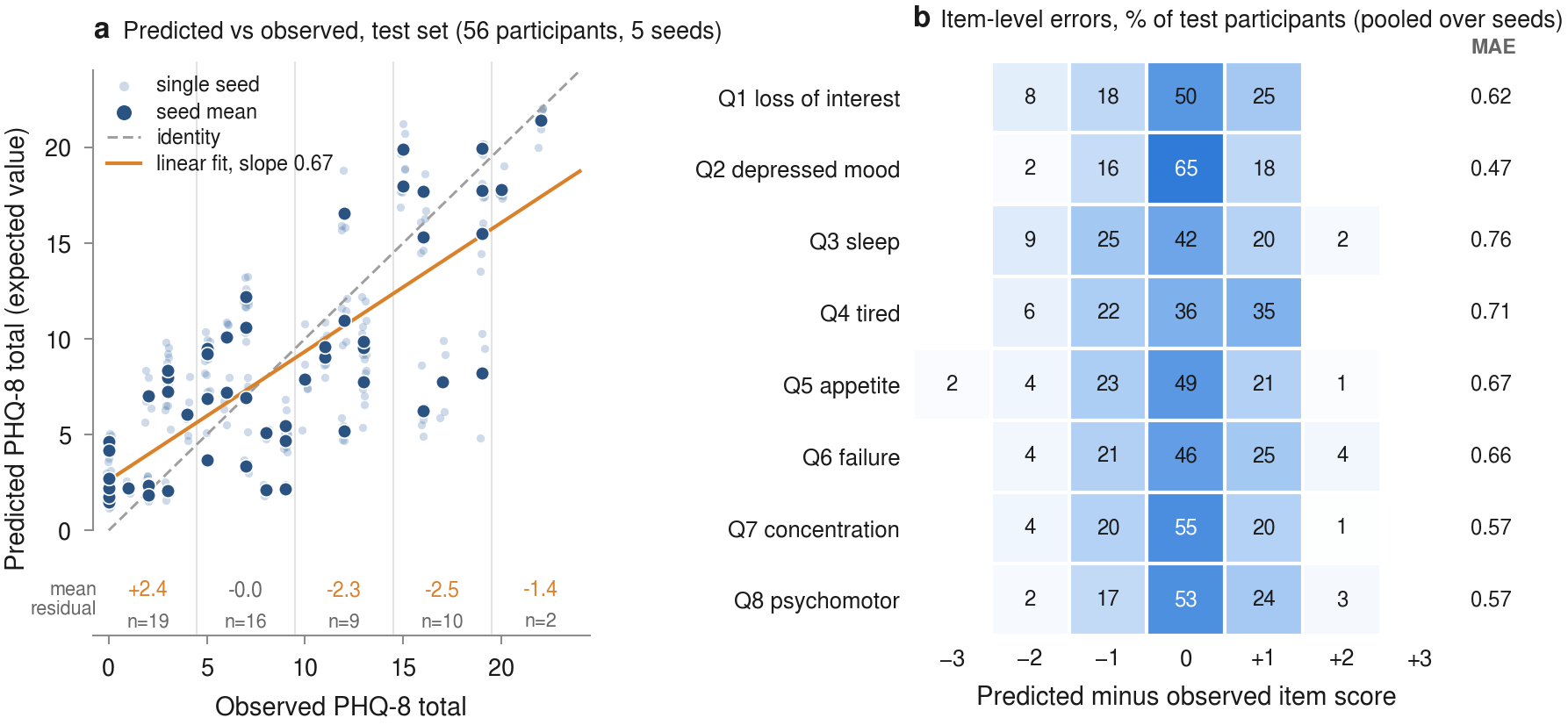}
\caption{\textbf{Predicted against observed severity for the trimodal SIGMA model on the test partition.} \textbf{a}, Predicted PHQ-8 total (expected-value decoding) against the observed total for the 56 test participants; large points are the mean over five seeds, small points the single seeds, the dashed line the identity and the orange line a linear fit. Below the axis, the mean residual (predicted minus observed) and the number of participants in each observed-score band. \textbf{b}, Item-level errors: for each PHQ-8 item, the percentage of test participants, pooled over the five seeds, whose rounded expected score differs from the observed category by the given amount; the right-hand column gives the per-item MAE of the expected score, averaged over seeds.}
\label{fig:predictions}
\end{figure}

\textbf{The improvement is uneven across severity levels and symptoms.} Where that gain sits determines what the model can be used for. Predictions track the observed totals ($r=0.78$) but over a compressed range: the slope of predicted on observed scores is 0.67 and the predictions have a standard deviation of 5.5 against 6.4 for the labels (Fig.~\ref{fig:predictions}a). Participants scoring 0--4 are over-predicted by 2.4 points on average, those scoring 10--19 are under-predicted by 2.3--2.5 points, and the largest errors fall in the 15--19 band (MAE 4.6, RMSE 5.9, $n=10$); only two test participants score 20 or above, so calibration at the severe end cannot be assessed on this partition. Seed-to-seed variation is small against these effects, with test CCC ranging from 0.731 to 0.780 over the five seeds.

The same unevenness appears across symptoms (Fig.~\ref{fig:predictions}b). The rounded expected score matches the observed category for 36\% (feeling tired) to 65\% (depressed mood) of participants and falls within one category for 88--98\% of them. Depressed mood, among the more skewed items, is predicted most accurately (MAE 0.47), whereas sleep problems and feeling tired, the most evenly distributed items, carry the largest errors (0.76 and 0.71), so per-item accuracy tracks label concentration rather than how directly a symptom is verbalised. Per-item results for every configuration and both partitions are given in Supplementary Tables~S2 and S3.

\begin{figure}[!ht]
\centering
\includegraphics[width=1.0\textwidth]{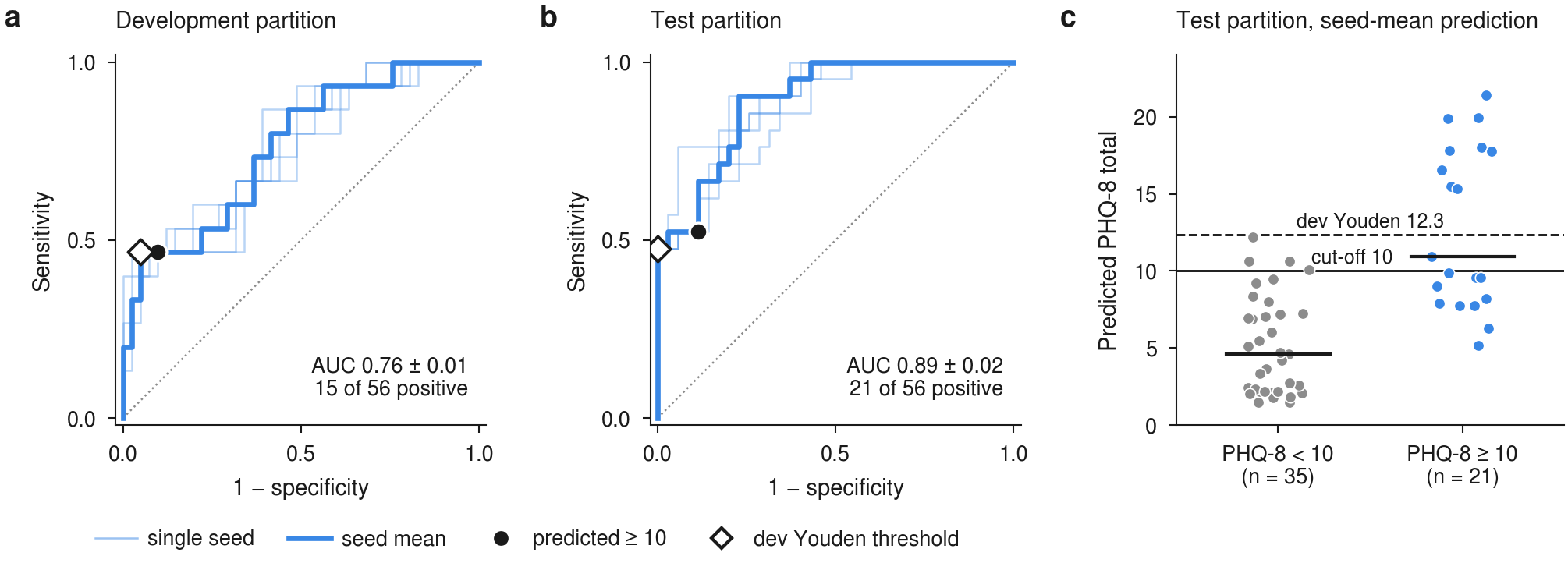}
\caption{\textbf{Screening performance of the trimodal SIGMA model at PHQ-8 $\geq 10$.} \textbf{a}, \textbf{b}, Receiver operating characteristic curves on the development (a) and test (b) partitions for the five seeds (thin) and their mean prediction (thick); AUC is mean $\pm$ s.d.\ over seeds. Filled circle: predicted total dichotomised at 10; open diamond: threshold maximising Youden's index on the development predictions of the seed-mean model. \textbf{c}, Seed-mean predicted totals of the test participants by observed screen status, with medians (horizontal bars), the cut-off of 10 (solid) and the development-selected threshold (dashed).}
\label{fig:screening}
\end{figure}

\textbf{Screening at the conventional cut-off favours specificity over sensitivity.} To evaluate the model as a screening test, we used the PHQ-8 threshold of 10 \cite{kroenke2009phq,wu2020equivalency}. The trimodal model classifies 15 of the 56 development and 21 of the 56 test participants as screen-positive.
The test partition yields an AUC of $0.89 \pm 0.02$, compared with $0.76 \pm 0.01$ on development (Fig.~\ref{fig:screening}a,b), mirroring the partition gap in CCC. 
At the same cut-off, test sensitivity is $0.59 \pm 0.10$ and specificity is $0.90 \pm 0.02$. The seed-mean predictions miss 10 of the 21 screen-positive test participants, including three with observed PHQ-8 scores of 16--19 (Fig.~\ref{fig:screening}c). This pattern is consistent with predictions being compressed towards the centre of the scale (Fig.~\ref{fig:screening}a). Selecting a threshold on the development partition does not improve performance, and its small sample size limits reliable threshold calibration.

In answer to RQ1, the framework improves on every published result available for comparison and in all three metrics at once, with residual error concentrated where the test partition provides fewest participants rather than spread evenly across the scale. For screening, SIGMA achieves a test AUC of 0.89, indicating good ranking ability, although sensitivity remains lower than specificity at the conventional cut-off. This accuracy was obtained with all three streams present, and RQ2 asks how much of it each stream is responsible for.

\begin{table}[!ht]
\centering
\caption{\textbf{Leave-one-out ablation of SIGMA.} Each row removes exactly one component from the full model, holding the input pipeline, partition, optimiser, schedule, checkpoint-selection rule and five random seeds fixed. Removed components are replaced by the alternative named in the text: speech isolation by the unprocessed streams, cross-attention by within-modality attention of matched depth and parameter count, the graded-response head by eight independent item heads, and joint training by eight single-item models per seed. Cross-attention is undefined for a single modality. Values are mean $\pm$ s.d.\ over five random seeds under expected-value decoding at the lowest-development-loss checkpoint. All values are on the official test partition; ``Item CCC'' is the mean per-item CCC. Boldface marks the best value within each modality block.}
\label{tab:our_results}
\footnotesize
\setlength{\tabcolsep}{3pt}
{\begin{tabular}{lcccc}
\hline
\textbf{Condition} & \textbf{CCC ($\uparrow$)} & \textbf{RMSE ($\downarrow$)} & \textbf{MAE ($\downarrow$)} & \textbf{Item CCC ($\uparrow$)} \\
\hline
\multicolumn{5}{l}{\textit{Text (T)}} \\
\quad Full model & 0.746\,\tiny{$\pm$\,0.039} & \textbf{4.08\,\tiny{$\pm$\,0.14}} & \textbf{3.36\,\tiny{$\pm$\,0.15}} & \textbf{0.544\,\tiny{$\pm$\,0.040}} \\
\quad $-$ speech isolation & 0.710\,\tiny{$\pm$\,0.038} & 4.38\,\tiny{$\pm$\,0.13} & 3.76\,\tiny{$\pm$\,0.07} & 0.519\,\tiny{$\pm$\,0.038} \\
\quad $-$ graded-response head & 0.727\,\tiny{$\pm$\,0.018} & 4.28\,\tiny{$\pm$\,0.07} & 3.48\,\tiny{$\pm$\,0.07} & 0.533\,\tiny{$\pm$\,0.023} \\
\quad $-$ joint training & \textbf{0.757\,\tiny{$\pm$\,0.018}} & 4.12\,\tiny{$\pm$\,0.12} & 3.39\,\tiny{$\pm$\,0.06} & 0.537\,\tiny{$\pm$\,0.016} \\
\hline
\multicolumn{5}{l}{\textit{Audio (A)}} \\
\quad Full model & 0.002\,\tiny{$\pm$\,0.003} & 6.40\,\tiny{$\pm$\,0.02} & \textbf{5.38\,\tiny{$\pm$\,0.02}} & 0.001\,\tiny{$\pm$\,0.002} \\
\quad $-$ speech isolation & 0.004\,\tiny{$\pm$\,0.007} & 6.40\,\tiny{$\pm$\,0.03} & 5.38\,\tiny{$\pm$\,0.02} & 0.003\,\tiny{$\pm$\,0.004} \\
\quad $-$ graded-response head & \textbf{0.014\,\tiny{$\pm$\,0.013}} & \textbf{6.37\,\tiny{$\pm$\,0.05}} & 5.38\,\tiny{$\pm$\,0.04} & \textbf{0.008\,\tiny{$\pm$\,0.008}} \\
\quad $-$ joint training & 0.003\,\tiny{$\pm$\,0.002} & 6.38\,\tiny{$\pm$\,0.01} & 5.40\,\tiny{$\pm$\,0.02} & 0.001\,\tiny{$\pm$\,0.003} \\
\hline
\multicolumn{5}{l}{\textit{Video (V)}} \\
\quad Full model & 0.016\,\tiny{$\pm$\,0.020} & 6.37\,\tiny{$\pm$\,0.03} & 5.39\,\tiny{$\pm$\,0.02} & 0.011\,\tiny{$\pm$\,0.013} \\
\quad $-$ speech isolation & \textbf{0.027\,\tiny{$\pm$\,0.030}} & \textbf{6.36\,\tiny{$\pm$\,0.01}} & \textbf{5.36\,\tiny{$\pm$\,0.09}} & \textbf{0.017\,\tiny{$\pm$\,0.018}} \\
\quad $-$ graded-response head & 0.000\,\tiny{$\pm$\,0.007} & 6.45\,\tiny{$\pm$\,0.04} & 5.40\,\tiny{$\pm$\,0.05} & 0.007\,\tiny{$\pm$\,0.005} \\
\quad $-$ joint training & -0.006\,\tiny{$\pm$\,0.003} & 6.41\,\tiny{$\pm$\,0.01} & 5.41\,\tiny{$\pm$\,0.02} & -0.005\,\tiny{$\pm$\,0.002} \\
\hline
\multicolumn{5}{l}{\textit{T + A}} \\
\quad Full model & \textbf{0.766\,\tiny{$\pm$\,0.019}} & 4.15\,\tiny{$\pm$\,0.18} & 3.31\,\tiny{$\pm$\,0.09} & \textbf{0.573\,\tiny{$\pm$\,0.013}} \\
\quad $-$ speech isolation & 0.677\,\tiny{$\pm$\,0.132} & 4.47\,\tiny{$\pm$\,0.36} & 3.83\,\tiny{$\pm$\,0.37} & 0.495\,\tiny{$\pm$\,0.116} \\
\quad $-$ cross-attention & 0.762\,\tiny{$\pm$\,0.022} & 4.14\,\tiny{$\pm$\,0.16} & \textbf{3.30\,\tiny{$\pm$\,0.06}} & 0.568\,\tiny{$\pm$\,0.020} \\
\quad $-$ graded-response head & 0.738\,\tiny{$\pm$\,0.010} & 4.35\,\tiny{$\pm$\,0.15} & 3.49\,\tiny{$\pm$\,0.12} & 0.548\,\tiny{$\pm$\,0.007} \\
\quad $-$ joint training & 0.757\,\tiny{$\pm$\,0.032} & \textbf{4.06\,\tiny{$\pm$\,0.11}} & 3.38\,\tiny{$\pm$\,0.13} & 0.524\,\tiny{$\pm$\,0.036} \\
\hline
\multicolumn{5}{l}{\textit{T + V}} \\
\quad Full model & 0.763\,\tiny{$\pm$\,0.020} & 4.14\,\tiny{$\pm$\,0.15} & 3.31\,\tiny{$\pm$\,0.08} & 0.569\,\tiny{$\pm$\,0.018} \\
\quad $-$ speech isolation & 0.695\,\tiny{$\pm$\,0.102} & 4.38\,\tiny{$\pm$\,0.33} & 3.78\,\tiny{$\pm$\,0.34} & 0.505\,\tiny{$\pm$\,0.089} \\
\quad $-$ cross-attention & \textbf{0.763\,\tiny{$\pm$\,0.019}} & 4.15\,\tiny{$\pm$\,0.14} & \textbf{3.30\,\tiny{$\pm$\,0.08}} & \textbf{0.570\,\tiny{$\pm$\,0.017}} \\
\quad $-$ graded-response head & 0.736\,\tiny{$\pm$\,0.010} & 4.37\,\tiny{$\pm$\,0.07} & 3.50\,\tiny{$\pm$\,0.08} & 0.547\,\tiny{$\pm$\,0.010} \\
\quad $-$ joint training & 0.747\,\tiny{$\pm$\,0.033} & \textbf{4.06\,\tiny{$\pm$\,0.15}} & 3.39\,\tiny{$\pm$\,0.18} & 0.502\,\tiny{$\pm$\,0.031} \\
\hline
\multicolumn{5}{l}{\textit{T + A + V}} \\
\quad Full model & \textbf{0.762\,\tiny{$\pm$\,0.021}} & 4.13\,\tiny{$\pm$\,0.17} & 3.30\,\tiny{$\pm$\,0.06} & \textbf{0.567\,\tiny{$\pm$\,0.017}} \\
\quad $-$ speech isolation & 0.735\,\tiny{$\pm$\,0.008} & 4.33\,\tiny{$\pm$\,0.06} & 3.67\,\tiny{$\pm$\,0.07} & 0.545\,\tiny{$\pm$\,0.012} \\
\quad $-$ cross-attention & 0.760\,\tiny{$\pm$\,0.021} & \textbf{4.10\,\tiny{$\pm$\,0.18}} & \textbf{3.27\,\tiny{$\pm$\,0.08}} & 0.563\,\tiny{$\pm$\,0.015} \\
\quad $-$ graded-response head & 0.743\,\tiny{$\pm$\,0.012} & 4.33\,\tiny{$\pm$\,0.14} & 3.48\,\tiny{$\pm$\,0.11} & 0.553\,\tiny{$\pm$\,0.009} \\
\quad $-$ joint training & 0.751\,\tiny{$\pm$\,0.008} & 4.17\,\tiny{$\pm$\,0.13} & 3.40\,\tiny{$\pm$\,0.08} & 0.528\,\tiny{$\pm$\,0.016} \\
\hline
\end{tabular}}
\end{table}

\subsection*{RQ2: Modality dependence and the limits of the non-verbal streams}
\label{sec:results-rq2}

\textbf{Accuracy survives the loss of audio and video but not of the transcript.} Clinical recordings do not always contain usable audio or video, so how much of the accuracy reported above survives their loss is a practical question as well as a scientific one. Comparing modality subsets of the complete framework, which differ only in the streams available to them, the decisive factor is not how many modalities are present but whether the transcript is among them (Table~\ref{tab:our_results}, full-model rows; development-partition CCC in Supplementary Table~S8). Text alone reaches CCC 0.746 with RMSE 4.08 and MAE 3.36 on the test partition, against CCC 0.002 for audio alone and 0.016 for video alone, whose RMSE of 6.37--6.40 is that of a model predicting close to the training mean; the development partition shows the same pattern (Supplementary Table~S8). Among the text-containing configurations the differences are small: a spread of 0.004 across the three multimodal models, and 0.019 once text alone is included, against a seed-to-seed standard deviation of up to 0.039.

\begin{figure}[!ht]
\centering
\includegraphics[width=1.0\textwidth]{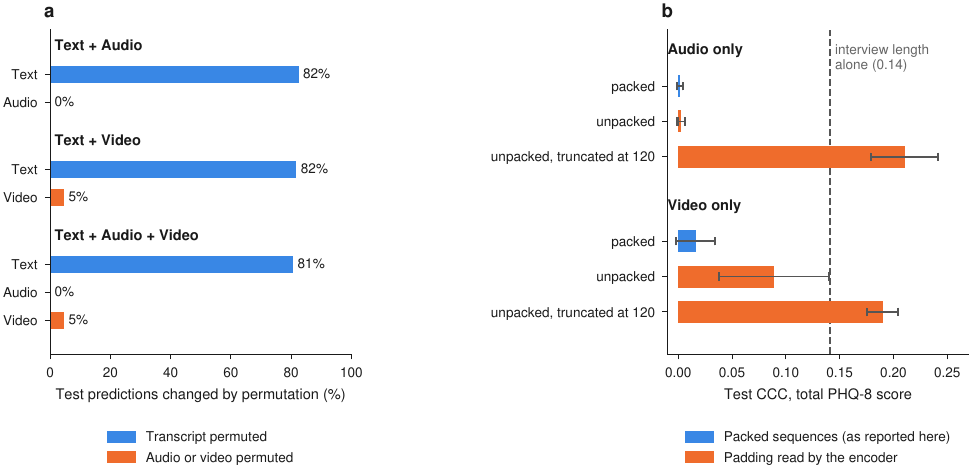}
\caption{\textbf{Permutation test  on contributions of acoustic and visual streams.} \textbf{a}, Permutation test on the released multimodal models: percentage of test predictions (argmax total) that change when the named modality's utterance sequence is permuted across participants, mean over five random seeds. Permuting the transcript changes most predictions; permuting audio or video changes almost none. \textbf{b}, Test CCC of the audio-only and video-only models under three sequence-handling conditions, mean over five random seeds with error bars giving the standard deviation: packed sequences, as reported throughout this paper; the same models with the recurrent encoder run over the padded tensor; and the padded tensor with sequences truncated at 120 utterances and eight independent item heads. The dashed line marks the test CCC of a linear predictor of the total score that uses the number of participant utterances alone. Apparent unimodal accuracy appears only when the encoder can read the padding, and then approaches the interview-length baseline. Numerical values are given in Supplementary Table~S7.}
\label{fig:nonverbal}
\end{figure}

\textbf{Permuting audio or video changes almost no predictions.} The small multimodal advantage over text alone raises a further question: does it come from non-verbal evidence, or from something else? A permutation test answers it, by breaking the correspondence between one stream and the participant and counting the predictions that change (Methods). Permuting the audio sequence changes none of the text--audio or trimodal predictions, permuting the video sequence changes 5\% of the text--video and of the trimodal predictions, and permuting the transcript changes 81--82\% (Fig.~\ref{fig:nonverbal}a). The trimodal and text-only predicted totals correlate at $r=0.995\pm0.004$ across seeds, and at 0.998 on the seed-mean predictions. This agreement is expected in part from the training procedure: each multimodal model starts from the trained text trunk of the same seed, with its weights frozen, so the transcript representation is inherited from the text-only model and only the audio and video encoders, the fusion and the head are trained further (Methods). With the utterance-level eGeMAPS and OpenFace descriptors distributed with the corpus, the non-verbal streams therefore appear to contribute little participant-specific evidence to the multimodal predictions, and the small multimodal advantage over text alone may reflect the two-stage training rather than what audio and video carry.

\textbf{Interview length inflates unimodal results reported without packing.} This leaves an apparent contradiction with the literature, in which audio-only and video-only models on E-DAIC reach non-trivial accuracy. Resolving it identifies a confound that extends beyond this framework. Sequences of unequal length are zero-padded to a common width before encoding. Unless the recurrent encoder is told where each sequence ends, it runs over the padded positions as well, and its backward state at the first real utterance then carries how much padding followed it, and with it the length of the interview (Methods). Interview length is not neutral on this corpus: a linear predictor using nothing but the number of participant utterances reaches CCC 0.141 on the test partition and 0.144 on the development partition (Supplementary Table~S7). Audio-only and video-only models that read the padding reach approximately 0.2 CCC through this route alone, against 0.002 and 0.016 for the packed models reported here (Fig.~\ref{fig:nonverbal}b). Unimodal accuracy on this corpus may therefore be driven in part by sequence handling rather than by acoustic or visual evidence. Every model reported here packs its sequences and excludes padded positions throughout.

In answer to RQ2, the framework degrades marginally as the non-verbal streams are removed and fails without the transcript. At this sample size and with the distributed utterance-level descriptors, audio and video appear to contribute little measurable participant-specific evidence, which likely bounds what any fusion mechanism could extract from them. The informative input is therefore the transcript, and RQ3 turns from the inputs to the stages that process them.

\subsection*{RQ3: Contribution of the four framework components}
\label{sec:results-rq3}

We answer RQ3 with the leave-one-out ablation specified in the Methods, in which each condition removes exactly one component and replaces it by a matched alternative, holding the input pipeline, partition, optimiser, schedule and random seeds fixed (Table~\ref{tab:our_results}; development-partition results in Supplementary Table~S8). The seed-to-seed standard deviation of CCC is 0.01--0.04 across the text-containing rows and sets the scale against which the differences below should be interpreted.

\textbf{Removing interviewer speech lowers accuracy in every configuration.} Restoring those turns is the most costly change in the ablation. Test CCC falls by 0.036 for text, 0.089 for text--audio, 0.068 for text--video and 0.027 for the trimodal model, while RMSE rises by 0.20--0.32 points and MAE by 0.38--0.52 points on every configuration; the development partition falls by a comparable margin (Table~\ref{tab:our_results} and Supplementary Table~S8). It is the only change in the study whose effect exceeds the seed-to-seed variation in all three total-score metrics. Its size varies with configuration: the penalty is 0.027--0.036 for text alone and the trimodal model and roughly doubles in the two bimodal configurations. It is also the only component that acts on the input rather than on the architecture. The audio-only and video-only models are at chance under both conditions, so the effect of isolation on the non-verbal streams cannot be measured here; for the visual stream, in which the participant remains on camera throughout, isolation changes only the utterance windows over which the OpenFace features are pooled, and those models do not differ between conditions.

\textbf{Cross-modal attention changes accuracy by less than the seed band.} Replacing it by within-modality attention of matched depth and parameter count lowers test CCC by 0.004 for text--audio and 0.002 for the trimodal model and leaves text--video unchanged, with differences of at most 0.005 at the item level and 0.002 on the development partition (Table~\ref{tab:our_results} and Supplementary Table~S8). RMSE and MAE differ by at most 0.03 points in either direction. All of these effects lie inside the seed band.

\textbf{Ordered outputs improve agreement and error on both partitions.} The graded-response head replaces eight free classifiers by one latent severity trait and eight item-specific response curves, so that the ordering of the four categories is a property of the output model rather than a preference expressed by the loss. Substituting eight independent item heads on the same fused representation lowers test CCC by 0.019 for text, 0.028 for text--audio, 0.027 for text--video and 0.019 for the trimodal model, lowers development CCC by a comparable margin, and raises RMSE by 0.20--0.23 points and MAE by 0.12--0.19 points on every text-containing configuration (Table~\ref{tab:our_results} and Supplementary Table~S8). Its effect is consistent in direction across every configuration and, with participant-speech isolation, it is one of two components whose removal worsens agreement and both error metrics together. The head has fewer free output parameters than eight independent classifiers, so the gain comes from the ordering and the shared severity scale it imposes rather than from any capacity it adds, and it also makes the model's account of each item readable, which RQ4 takes up.

\textbf{Joint training improves item-level agreement at a fraction of the cost.} Sharing one trunk across the eight items raises the mean per-item CCC from 0.502--0.537 to 0.544--0.573, a gain of 0.007 for text and 0.039--0.067 for the three multimodal configurations, and the largest item-level effect of any component apart from speech isolation (Table~\ref{tab:our_results}). On the total score the two designs are close, the shared trunk giving the higher CCC on every multimodal configuration and the single-item models the lower RMSE in the two bimodal ones (4.06 against 4.14--4.15). Joint training therefore improves the item-level predictions the framework is designed to produce while leaving the total score essentially unchanged, and does so at a fraction of the cost, since the eight single-item models require eight separate trunks per seed.

In answer to RQ3, three of the four components contribute measurably, each under the conditions its mechanism addresses: participant-speech isolation improves what the model sees, the graded-response head improves how it reports, and joint training improves the item-level output the framework exists to produce. Cross-modal attention does not, consistent with RQ2: there is little non-verbal evidence for it to exploit.

\subsection*{RQ4: What the model learns about the individual symptoms}
\label{sec:results-rq4}

\begin{figure}[!ht]
\centering
\includegraphics[width=1.0\textwidth]{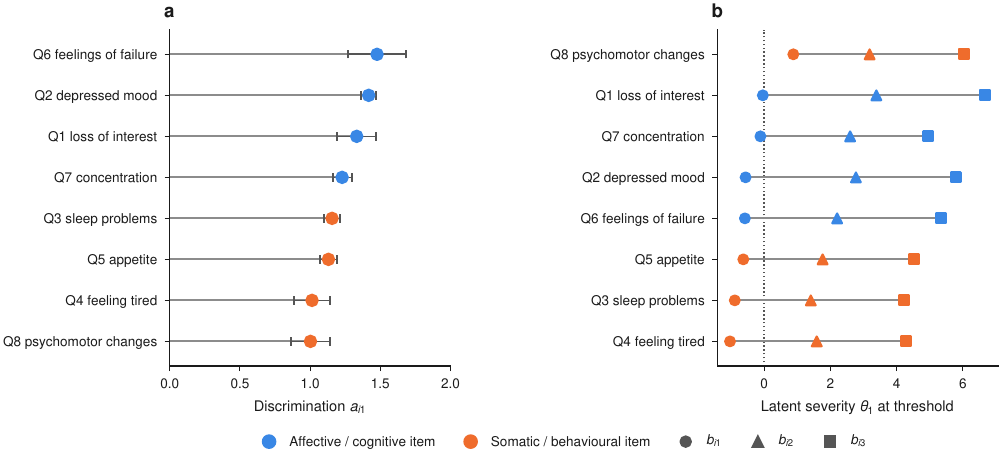}
\caption{\textbf{Fitted item parameters of the graded-response head.} Trimodal SIGMA model, first latent factor, mean over five random seeds. Items are coloured by symptom type: affective and cognitive items (Q1, Q2, Q6, Q7) and somatic and behavioural items (Q3, Q4, Q5, Q8). \textbf{a}, Discrimination $a_{i1}$, the loading of each item on the latent severity trait, ordered by magnitude; error bars are the standard deviation over the five seeds. The four affective and cognitive items rank above all four somatic and behavioural items. \textbf{b}, The three ordered thresholds $b_{i1}<b_{i2}<b_{i3}$, the trait values at which the probability of scoring at least 1, 2 and 3 on the item reaches one half, ordered by $b_{i1}$. Items to the left are endorsed at mild severity and items to the right only at severe levels. Numerical values are given in Supplementary Table~S9 and the corresponding response curves in Supplementary Fig.~S1.}
\label{fig:item_params}
\end{figure}

Since the graded-response head places all eight items on one severity scale, each item is summarised by two quantities that can be interpreted directly: a discrimination, how sharply the item responds as severity rises, and three thresholds, the severity at which each successive category becomes the more likely answer. Both are free parameters learned with the rest of the model, initialised from the category frequencies of the training partition (Methods). Eight independent classifiers permit no such interpretation, since each estimates four category scores carrying neither an ordering nor a scale shared with the other items. Both quantities are stable across seeds (discrimination s.d.\ 0.05--0.21; Fig.~\ref{fig:item_params}, Supplementary Table~S9 and Supplementary Fig.~S1). They describe the model's own item behaviour rather than the psychometrics of the PHQ-8 in this population.

\textbf{Affective symptoms discriminate more sharply than somatic ones.} The four affective and cognitive items rank above all four somatic and behavioural items without exception (Fig.~\ref{fig:item_params}a): feelings of failure ($a_{i1}=1.48$), depressed mood (1.42), loss of interest (1.33) and concentration (1.23) above sleep problems (1.16), appetite (1.13), feeling tired (1.01) and psychomotor changes (1.00). Since the model reads only what participants say, this ordering reflects how far each symptom is described in the interview rather than the psychometric quality of the item. It is therefore a statement about the interview as a measurement channel: the symptoms that most sharply locate a participant on the severity scale are those put into words, so recovering the somatic items would require explicit questioning or an observational channel rather than better modelling of the same transcript.

\textbf{Thresholds follow the order in which symptoms are endorsed.} Feeling tired ($b_{i1}=-1.03$) and sleep problems ($-0.89$) have the lowest thresholds and are endorsed at mild severity, whereas psychomotor changes ($+0.88$) and loss of interest ($-0.04$) have the highest and appear only at more severe levels (Fig.~\ref{fig:item_params}b). This ordering was not supplied to the model.

Together, the two parameters indicate which symptoms are informative, and when. Feeling tired and sleep problems are endorsed at the mildest severities but discriminate least, so they signal that a participant is distressed without indicating how severely. Feelings of failure and depressed mood discriminate most sharply from intermediate thresholds, and therefore carry most of the information about where on the scale a participant falls. Loss of interest is endorsed only at higher severity and still discriminates sharply, making it both specific to more severe presentations and informative within them, whereas psychomotor change is endorsed latest of all and discriminates least. A total score collapses these distinctions; estimating the two parameters per item is what recovers them.

In answer to RQ4, the graded-response head yields item parameters that are stable across seeds and directly readable: discriminations that separate the affective and cognitive symptoms from the somatic ones, and thresholds that reproduce the order in which symptoms are endorsed as severity rises.

\section*{Discussion}
\label{sec:discussion}
\addcontentsline{toc}{section}{Discussion}

In this paper, we studied the automated estimation of depression severity from semi-clinical interviews. We developed SIGMA, a framework that isolates participant speech, allows text, audio and video to condition one another during encoding, and predicts the eight ordinal PHQ-8 items jointly through a graded-response head. On E-DAIC the trimodal model achieved a CCC of 0.762, an RMSE of 4.13 and an MAE of 3.30 on the official test partition, averaged over five random seeds, outperforming published results in our comparison. A leave-one-out ablation under a matched protocol indicates that the four components of the framework contribute in different places, and each is considered below alongside the prior work it builds on.

Speaker identity has received little attention in depression-detection research. It is irrelevant to the social-media text on which early work was built, and the shift to semi-clinical corpora introduced a second speaker whose prompts, prosody and turn-taking are mixed with the participant's. Speaker diarization is a well-established remedy \cite{khoma2023development}, but it has rarely been evaluated inside a multimodal severity-prediction pipeline. 
This issue extends beyond depression detection. Analyses of recorded interviews, from automated assessment of other mental-health conditions to computational studies of qualitative research interviews, risk attributing the interviewer’s language, prosody, or questioning style to the person being studied. Furthermore, the proportion of interviewer speech may vary systematically across sessions, sites, or data partitions, as observed in this study. Speaker separation may therefore be an important consideration in interview-based research more broadly, rather than a concern limited to this particular task.
Our ablation isolates participant-speech isolation while holding the architecture, features, training procedure, and seeds fixed. It yields the largest gains, improving agreement and both error metrics across all text-containing configurations. As an input-level intervention, it offers gains independent of architectural choices. However, diarization accuracy on E-DAIC remains unverified, as participant identity is assigned using an unvalidated dominant-speaker heuristic.

Multimodal depression modelling has moved from feature concatenation and independently encoded streams \cite{nasir2016multimodal,alhanai2018detecting,rodrigues2019multimodal} to cross-modal attention, adaptive fusion and graph-based representations \cite{tsai2019multimodal,sun2021multi,yuan2024depression,shen2024multi}. These generally outperform unimodal baselines, but they apply one fusion strategy across the whole interview. Against that trend, Sadeghi et al.\ \cite{sadeghi2024harnessing} obtained the lowest published text--video errors by extracting depression indicators from the transcripts with a proprietary large language model and combining them with facial-behaviour features and speech-quality filtering, which suggests that preparing the inputs may matter as much as the sophistication of the fusion. SIGMA improves on those errors with openly available encoders run locally, so the accuracy reported here does not depend on access to a commercial model. Our ablation supports this interpretation: the largest improvement comes from changing what is presented to the model rather than how the modalities are combined, and cross-modal attention changes agreement by less than the seed-to-seed variation.

Whether that margin is a limit of the mechanism or of the data cannot be settled here. Audio and video should carry paralinguistic and behavioural evidence that text does not \cite{li2024fpt}, yet with the utterance-level OpenSMILE and OpenFace descriptors distributed with E-DAIC \cite{stepanov2018depression,xu2023multimodal,zhao2020hierarchical} the transcript carries the predictive signal. The ablation establishes that fusion is not the bottleneck, since every variant tested was limited by the information available in the streams rather than by its design, and the attention diagnostics concur: in trained models the attention weights are uniform over the utterances, so each cross-modal module reduces to mean pooling, which is little more than a stream of utterance means can offer. End-to-end acoustic and visual encoders, and features that retain within-utterance dynamics, would separate the two possibilities. Two consequences follow: unimodal audio and video baselines on this corpus should be reported with packed sequences or alongside an interview-length baseline, since an unpacked recurrent encoder can read interview length out of the padding; and a deployment that loses the camera or the microphone retains essentially all of the framework's accuracy.

The PHQ-8 motivates moving beyond total-score regression, since the total aggregates eight distinct but related ordinal symptoms \cite{kroenke2009phq}. Van Steijn et al.\ \cite{van2022text} predicted individual items for interpretability, Milintsevich et al.\ \cite{milintsevich2023towards} encoded each conversational turn to capture turn-level evidence, and Mandal et al.\ \cite{mandal2025enhancing} introduced question-wise fusion; the ordinal and imbalanced distribution of item scores has in turn motivated objectives derived from ordinal log-loss \cite{castagnos2022simple,aurelio2019learning}. These approaches recognise item-level structure but leave the ordering of the four categories to the loss rather than to the model.

The graded-response head encodes that ordering in the model, as a latent severity trait with per-item response curves that are ordered by construction. After participant-speech isolation, which acts on the input, it is the component with the most consistent effect: removing it worsens agreement and both error metrics on every text-containing configuration and on both partitions, and it achieves this with fewer free output parameters than eight independent classifiers. Multi-task learning is complementary, preserving item-specific outputs while sharing information across related tasks \cite{zhang2018overview,ruder2017overview}: sharing one trunk across the eight items raises per-item agreement on every configuration and leaves the total score essentially unchanged against eight single-item models, at a fraction of their cost. The items are dominated by a single common factor (Cronbach's $\alpha=0.911$; Supplementary Table~S1), which is what the graded-response head encodes, and the shared trunk adds the item-level precision the framework exists to deliver. This suggests that item-response modelling, trained jointly, is an appropriate way to encode symptom structure at this scale. The findings are consistent with international psychometric evaluation studies for the PHQ-8, which map responses onto a single underlying factor \cite{de2023reliability}.

The fitted parameters characterise the measurement channel as well as the model. Since SIGMA reads only what participants say, an item's discrimination reflects how far the symptom is described in the interview, and on this scale every affective and cognitive item ranks above every somatic and behavioural one. Published item-response analyses of the PHQ-9 agree in part: Gao and Liu \cite{gao2024phq9irt} also report depressed mood as the most discriminating item, but place psychomotor change among the more discriminating ones, where it is the least discriminating here. Psychomotor change is the symptom a clinician would normally observe rather than be told about, so the discrepancy is what a transcript-only input would predict. The fitted thresholds, by contrast, reproduce the clinical ordering in which sleep and energy disturbance are endorsed before psychomotor change, without that ordering being supplied to the model. Item-level output of this kind renders a severity estimate inspectable rather than merely numerical.

Several limitations constrain these findings. E-DAIC contains 275 sessions and 56 test participants, and the test partition has a higher label mean and variance than those of the training and development partitions.
The results may therefore be sensitive to sampling variation. Although performance is averaged over five seeds, the limited sample size constrains the reliability of confidence intervals and participant-level significance tests, and small differences between configurations should be interpreted cautiously. 
Also, the participants are English-speaking adults in the United States recruited from the general and veteran populations rather than from clinical services, and they were interviewed by a virtual agent rather than a clinician.
The extent to which these findings generalise to clinical populations, other languages and cultural contexts, and clinician-led interviews remains to be established.
Comparisons with published methods warrant the same care, since they differ in feature extraction, in whether item-level or total scores are modelled and in whether results are averaged over seeds. We report the development partition alongside the test partition throughout, while noting that it also selects checkpoints and loss hyperparameters and is therefore not an independent held-out estimate. Identifying the participant as the dominant diarized speaker may fail where interviewer turns are long or speech overlaps, and these assignments were not manually validated. Finally, the model predicts questionnaire scores on a benchmark corpus and has not been evaluated prospectively, across clinical sites, or for demographic fairness; it should not be interpreted as a diagnostic system.

In summary, SIGMA estimates PHQ-8 severity at the item level by isolating participant speech, letting text, audio and video condition one another during encoding, and predicting the eight items jointly through a graded-response head. A leave-one-out ablation under a matched protocol shows that speech isolation, the graded-response head and joint training each contribute, while the benefit of cross-modal fusion is bounded by the non-verbal features available on this corpus. 
The fitted item parameters are stable across seeds and clinically interpretable. Since the predictions compress towards the mean, the nearer-term use is screening, where the decision is whether to refer rather than what the exact score is, rather than tracking change within a participant, and in either case the aim is to support clinical judgement rather than to replace it. Future work should validate the framework on larger and more diverse clinical datasets, incorporate diarization confidence into downstream modelling, evaluate richer non-verbal representations and forms of symptom interaction, extend the approach to related conditions such as anxiety and post-traumatic stress, and examine longitudinal symptom trajectories across repeated sessions.

\section*{Methods}
\label{sec:method}
\addcontentsline{toc}{section}{Methods}

This section describes the dataset, the prediction task, the four components of the framework, and the experimental procedures summarised in the Results.

\subsection*{E-DAIC dataset}
\label{sec:method-dataset}

E-DAIC \cite{devault2014simsensei} extends DAIC-WOZ \cite{gratch2014distress} and is a widely used benchmark for automated depression assessment. Participants were 275 adults in the United States recruited from the general population and a veterans’ population through online advertisements and flyers, rather than selected by clinical diagnosis. The corpus was designed to support the assessment of psychological distress, including depression and post-traumatic stress disorder (PTSD), and thus covers the full range of PHQ-8 severity rather than a clinically selected sample.

Participants completed the PHQ-8 and PCL-C before a semi-structured interview with Ellie, a virtual interviewer operating in either human-controlled Wizard-of-Oz or fully autonomous mode. The test partition contains only autonomous-mode sessions, whereas the training and development partitions include both modes \cite{sadeghi2023exploring,sadeghi2024harnessing}. Alongside the audio recordings and interview transcripts, the corpus provides pre-extracted acoustic features from OpenSMILE \cite{eyben2010opensmile} and facial action units, head pose and gaze direction from OpenFace \cite{baltruvsaitis20182}. Each session includes PCL-C and PHQ-8 annotations; only the PHQ-8 is used here. Supplementary Table~S1 reports partition sizes and total-score statistics; the score distribution is right-skewed, which motivates the imbalance-aware objective below.

\subsection*{Depression-severity prediction task}
\label{sec:method-problem}

We formulate depression assessment as an item-level multimodal prediction task. We represent an interview as an utterance sequence $\mathcal{U}=\{u_1,\dots,u_L\}$, where utterance $u_i$ has transcript text $t_i$, acoustic features $a_i$ and visual features $v_i$. The target is the item-level severity vector $\mathbf{y}=(y_1,\dots,y_8)\in\{0,1,2,3\}^8$. We sum the eight predicted item scores to obtain the total PHQ-8 score in $[0,24]$.

\subsection*{Participant-speech isolation and multimodal alignment}
\label{sec:method-diarization}

\begin{figure}[!ht]
\centering
\includegraphics[width=1.0\textwidth]{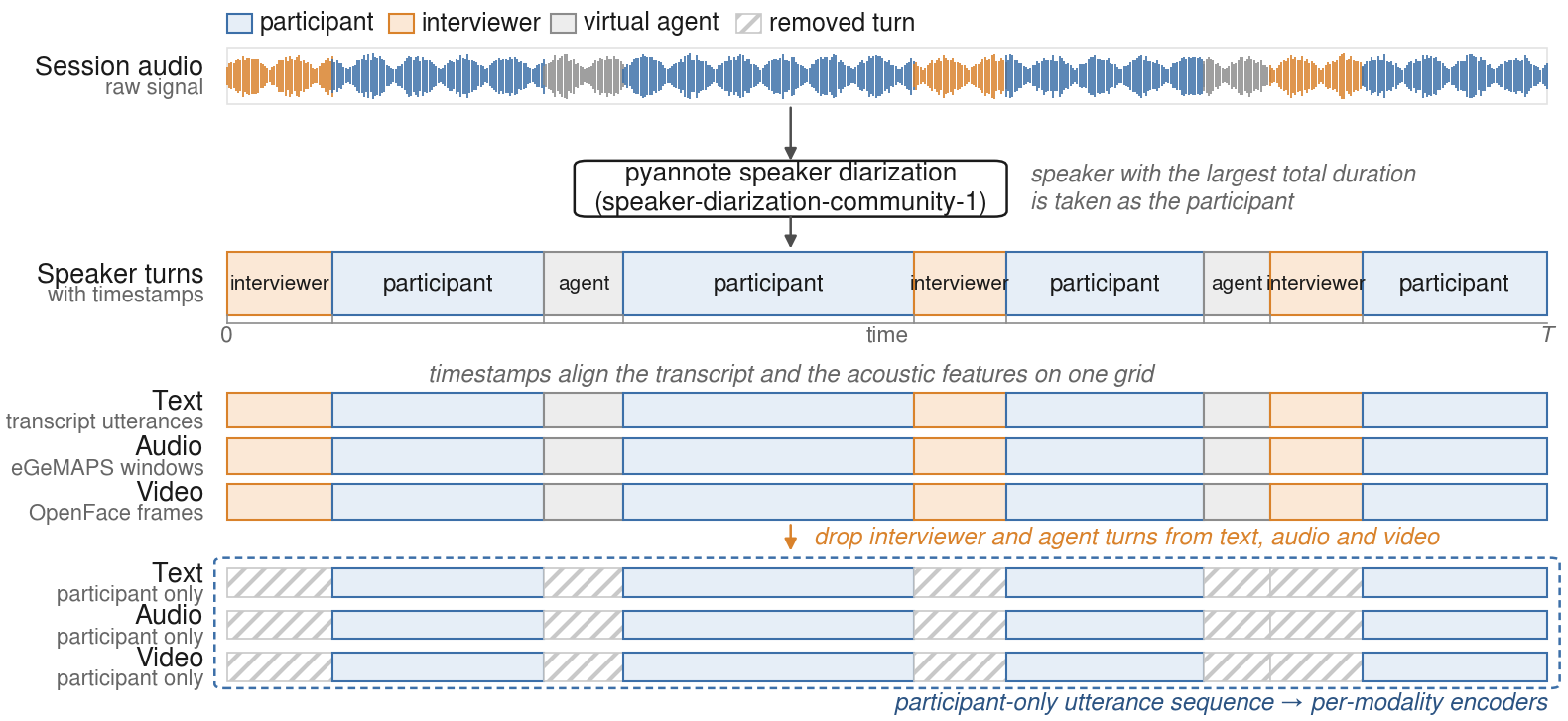}
\caption{\textbf{Speaker diarization and multimodal alignment.} The pretrained \emph{speaker-diarization-community-1} model from \emph{pyannote} \cite{bredin2020pyannote,Plaquet23} is applied to the raw audio of each E-DAIC session to identify speaker turns and produce timestamps for every turn. The resulting speaker labels are aligned with the transcript and the acoustic features on a common temporal grid. Interviewer and virtual-agent turns are discarded and participant turns are retained, yielding a participant-focused multimodal sequence that is consumed by the fusion stage. The visual stream is pooled over the same participant turns.}
\label{fig:Pyannote_process}
\end{figure}

Clinical interview recordings contain both participant and interviewer speech, whereas the prediction target describes the participant. We therefore identified participant speech segments before extracting downstream representations. We applied the pretrained \texttt{speaker-diarization-\allowbreak community-1} model from pyannote \cite{bredin2020pyannote,Plaquet23} to the original interview audio; the model assigns a speaker label and start and end timestamps to each dialogue turn. Since participants typically contribute the largest proportion of speech in each session, while the interviewer mainly provides shorter prompts and questions, we assigned the speaker with the greatest total speaking duration to the participant and retained the corresponding segments. We applied this heuristic to all sessions. It may fail when interviewer speech dominates or when diarization errors occur, and the speaker assignments were not manually validated. The retained timestamps select the transcript utterances and define the windows over which the OpenSMILE and OpenFace features are pooled, producing participant-focused text, audio and video streams (Fig.~\ref{fig:Pyannote_process}).

After diarization, we synchronised the three modalities at the transcript-utterance level to obtain a common temporal unit before encoding. We encoded each utterance with the pretrained sentence encoder \texttt{bge-large-en-v1.5} \cite{xiao2024cpack}, taking the L2-normalised CLS embedding to give a text embedding of dimension $d_t$; this encoder was selected among six candidates by five-fold cross-validation over all participants, which gave it the lowest cross-validated error (Supplementary Table~S6), and it was not fine-tuned. For audio and video, we mean-pooled the frame-level OpenSMILE eGeMAPS and OpenFace features, of dimensions $d_a$ and $d_v$ respectively, over the time interval of each transcript utterance, and L2-normalised the video vector. In 83 of the 275 original transcripts the last row carries the session's first start time instead of its own, so that its window spans the whole interview; we estimated its start from the utterance length (0.45 s per word and at least 1 s, not earlier than the previous end) and applied this window repair in every configuration. This yielded one textual, acoustic and visual vector per participant utterance. As sessions vary in the number of participant utterances, each modality-specific sequence is zero-padded to a common length $T_{\max}$, the longest original transcript in the corpus (203 utterances), and packed, so that no interview is truncated and the recurrent encoders process only the real utterances. Each participant is therefore represented by three temporally synchronised sequences of equal length before fusion.

\subsection*{Multimodal cross-attention fusion}
\label{sec:method-fusion}

\begin{figure}[!ht]
\centering
\includegraphics[width=1.0\textwidth]{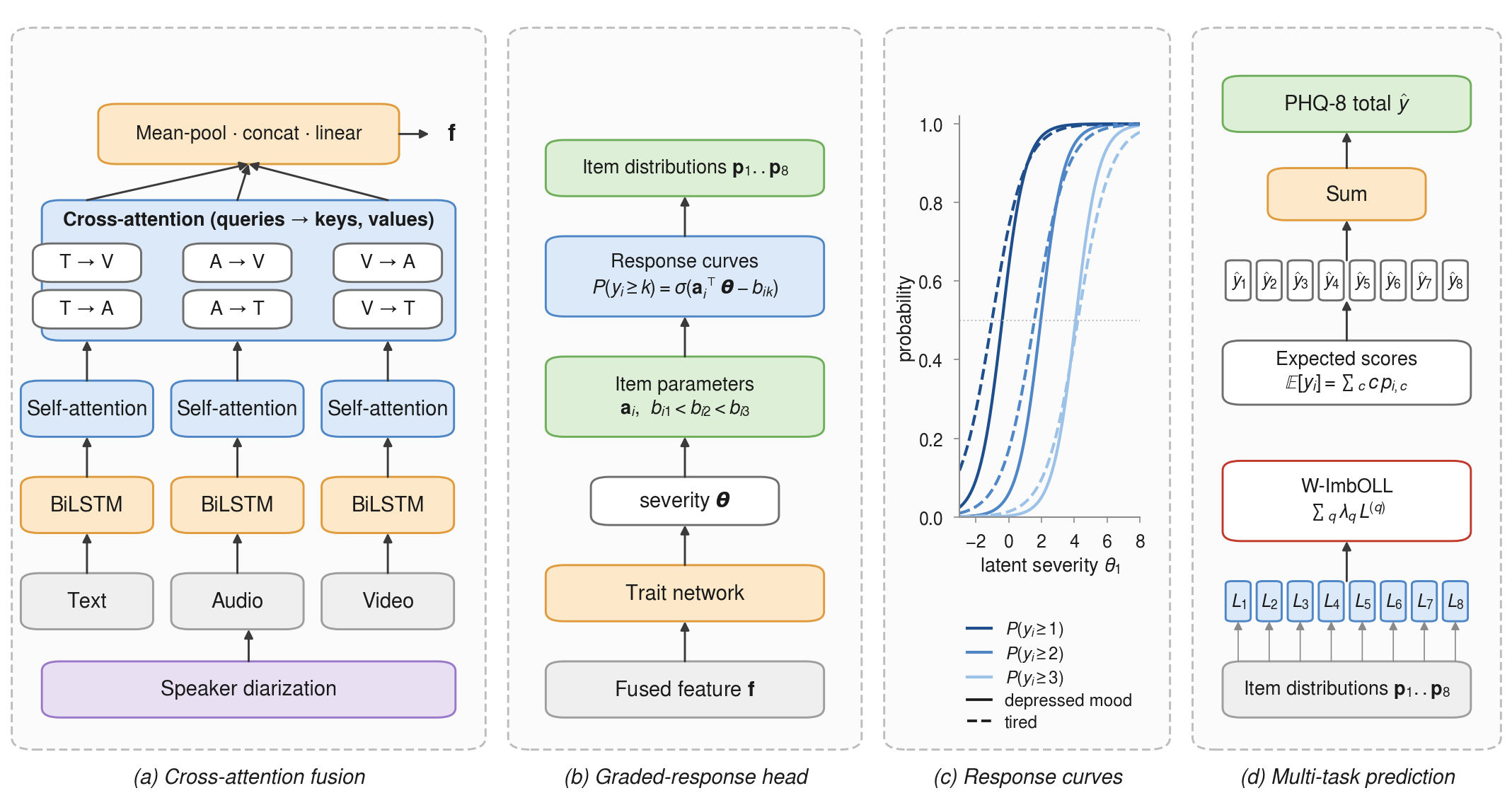}
\caption{\textbf{SIGMA architecture.} \textbf{a}, Multimodal cross-attention fusion. Participant-only text, audio and video sequences from the upstream diarization stage are encoded by per-modality BiLSTM and self-attention towers; their outputs are routed through pairwise cross-attention (all six ordered pairs in the trimodal model), mean-pooled over the real utterances, then concatenated and linearly projected to a fused participant representation $\mathbf{f}$. \textbf{b}, Graded-response head. A small network maps $\mathbf{f}$ to a latent severity trait $\boldsymbol{\theta}$; each item has a learned discrimination vector $\mathbf{a}_i$ and three ordered thresholds $b_{i1}<b_{i2}<b_{i3}$, and the probability of scoring at least $k$ on item $i$ is $\sigma(\mathbf{a}_i^{\top}\boldsymbol{\theta}-b_{ik})$. \textbf{c}, Fitted response curves of the trimodal model on the first factor for two items, depressed mood (solid) and feeling tired (dashed): the cumulative probabilities $P(y_i\geq1)$, $P(y_i\geq2)$ and $P(y_i\geq3)$ as functions of the latent severity, whose steepness is the item's discrimination and whose crossings of one half are its thresholds. \textbf{d}, Multi-task severity prediction. The eight item distributions are trained jointly with a Weighted Imbalanced Ordinal Log-Loss with learned per-item weights $\lambda_q$, and the total PHQ-8 score is the sum of the expected item scores.}
\label{fig:The_Proposed_Multi-modal_Multi-task_Architecture}
\end{figure}

Before fusion, three modality-specific single-layer BiLSTMs process the text, audio and visual feature sequences independently and capture within-modality temporal dependencies. Self-attention layers \cite{vaswani2017attention} then capture longer-range contextual relationships in each sequence, producing modality-specific representations $\mathbf{H}_t, \mathbf{H}_a, \mathbf{H}_v \in \mathbb{R}^{T_{\max} \times d}$. Each BiLSTM has $h$ hidden units per direction, giving concatenated states of dimension $d=2h$, and each self-attention layer has $H$ heads, per-head key dimension $d_k=d/H$ and dropout rate $p$. The BiLSTM and self-attention parameters are not shared across modalities, allowing each stream to learn patterns specific to its modality (Fig.~\ref{fig:The_Proposed_Multi-modal_Multi-task_Architecture}a).

To model cross-modal interactions \cite{tsai2019multimodal}, we apply pairwise cross-attention modules. For an ordered pair $\mathbf{H}_x \rightarrow \mathbf{H}_y$, we project queries from $\mathbf{H}_x$ and keys and values from $\mathbf{H}_y$, producing a sequence in $\mathbb{R}^{T_{\max} \times d}$ in which the utterances of modality $x$ select information from modality $y$. Each module uses $H$ heads, per-head key dimension $d_k$ and dropout $p$. The trimodal model uses all six ordered pairs, so that each modality supplies queries against each of the other two and keys and values to each of them, and the bimodal models use the two streams of their pair. Each stream is mean-pooled over the real utterances of its query modality, excluding padded positions, and the pooled vectors, six for the trimodal model and two for a bimodal one, are concatenated and linearly projected to the unified participant representation $\mathbf{f} \in \mathbb{R}^{d_f}$; the self-attention outputs $\mathbf{H}_t$, $\mathbf{H}_a$ and $\mathbf{H}_v$ enter $\mathbf{f}$ only through the cross-attention modules. A unimodal model has no cross-attention: its participant representation $\mathbf{f}$ is the masked mean of its self-attention output, of dimension $d$, used without projection.

\subsection*{Graded-response head for item-level ordinal labels}
\label{sec:method-grm}

The four PHQ-8 response categories, ``not at all'', ``several days'', ``more than half the days'' and ``nearly every day'', are ordered but not interval-scaled: the increase in severity from ``more than half the days'' to ``nearly every day'' need not equal that from ``not at all'' to ``several days''. A classifier over four nominal categories represents neither property. An ordinal loss represents the ordering, but the categories reach it as the integers 0 to 3, so their spacing remains an assumption rather than a quantity the model can estimate. We therefore replace independent per-item classifiers by a graded-response model \cite{samejima1969estimation}, the item-response formulation used in psychometrics for ordered polytomous items, in which the four categories are regions of a continuous severity scale whose boundaries are learned from the data.

Specifically, a trait network, a three-layer perceptron with dimensions $d_f\rightarrow d_1\rightarrow d_2\rightarrow K$ with rectified linear activations and dropout (0.3 after the first and 0.2 after the second hidden layer), maps the unified participant representation $\mathbf{f}$ to a latent severity trait $\boldsymbol{\theta}\in\mathbb{R}^{K}$. In the unimodal models the trait network has one hidden layer of 256 units, with dropout 0.5 on its input (0.2 for video) and 0.2 after the hidden layer; its first layer reads the pooled vector tiled 120 times through a flatten layer, a form kept from the reference implementation, whose flatten layer read a 120-utterance sequence, and equivalent to a linear layer on the pooled vector (Supplementary Table~S4). Each item $i$ has a learned discrimination vector $\mathbf{a}_i\in\mathbb{R}^{K}$, whose first component is constrained positive through a softplus, and three ordered thresholds $b_{i1}<b_{i2}<b_{i3}$, parameterised as a free $b_{i1}$ and softplus-positive increments. The probability of scoring at least $k$ on item $i$ is
\begin{equation}
P\!\left(y_i \geq k \mid \mathbf{f}\right) = \sigma\!\left(\mathbf{a}_i^{\top}\boldsymbol{\theta} - b_{ik}\right), \qquad k\in\{1,2,3\},
\end{equation}
and the class probabilities are the successive differences $p_{i,0}=1-P(y_i\geq1)$, $p_{i,k}=P(y_i\geq k)-P(y_i\geq k+1)$ and $p_{i,3}=P(y_i\geq3)$, which are ordered by construction. The ordering of the categories therefore holds in the output model rather than only in the loss; the discriminations state how strongly each item responds to severity and the thresholds where its response steps occur, and both are reported (Fig.~\ref{fig:item_params} and Supplementary Table~S9). The thresholds are initialised from the training marginals, so that at $\boldsymbol{\theta}=\mathbf{0}$ the cumulative probabilities equal the empirical frequencies, and the item parameters are trained at fifty times the base learning rate without weight decay; without either, they remain at their initial values.

\subsection*{Multi-task severity prediction and training objective}
\label{sec:method-prediction}

Unlike methods that predict only the total PHQ-8 score or model each item independently \cite{mandal2025enhancing,sadeghi2024harnessing}, we formulate item-level severity estimation as a multi-task learning problem with eight related tasks that share the trunk, the fusion and the trait network and differ only in their item parameters, which enables shared representation learning while retaining item-specific outputs \cite{zhang2018overview,ruder2017overview}. At inference the total PHQ-8 estimate is the sum of the expected item scores, $\sum_i\sum_c c\,p_{i,c}$, which is the natural output of the graded-response model; the sum of the per-item modal classes, an integer on $[0,24]$, is reported for comparison in the Results, since most published E-DAIC systems output integer scores.

The eight tasks are trained jointly with a Weighted Imbalanced Ordinal Log-Loss (W-ImbOLL). This builds on the Ordinal Log-Loss \cite{castagnos2022simple,aurelio2019learning} and the Imbalanced Ordinal Log-Loss \cite{mandal2025enhancing} by incorporating both ordinal error distance and class imbalance. The loss and the graded-response head address the ordinal structure at two complementary points. W-ImbOLL acts during training, on whatever distribution the model produces, and encodes the ordering by counting the categories that separate a prediction from the true response: answering ``nearly every day'' when the true response is ``not at all'' is three categories out and is penalised three times as heavily as answering ``several days'', which is one category out. Counting categories, however, makes every step the same size, so the loss also asserts that moving from ``not at all'' to ``several days'' is exactly as large a change in severity as moving from ``several days'' to ``more than half the days''. The graded-response head acts on the output space instead: it constrains the category probabilities to be ordered by construction and places the four categories on a latent severity scale at thresholds estimated from the data, so that the first of those steps may be learned as larger or smaller than the second rather than fixed in advance. For four ordinal classes, the item-level loss is
\begin{equation}
\mathcal{L}_{\text{ImbOLL}}^{(q)}
=-\frac{1}{N}\sum_{n=1}^{N}\sum_{c=0}^{3}
\log\!\big(1-p_{n,c}^{(q)}\big)\,
w\!\big(y_n^{(q)}\big)^{\beta}\left|y_n^{(q)}-c\right|^{\alpha},
\end{equation}
where $p_{n,c}^{(q)}$ is the predicted probability of class $c$ for sample $n$ and item $q$, and $y_n^{(q)}$ is the ground-truth class. The distance term penalises probability assigned to classes farther from the ground truth, so that larger errors across severity levels incur greater penalties. The inverse-frequency weight is $w(y_n^{(q)})=N/N_{y_n}^{(q)}$, where $N_{y_n}^{(q)}$ is the number of training samples of that class for item $q$, giving higher weight to under-represented severity levels. The multimodal models use these weights as fixed constants; the unimodal models initialise them at the same values and learn them with the rest of the parameters, kept positive through a softplus (Supplementary Table~S4). The hyperparameters $\alpha$ and $\beta$ control the ordinal-distance and class-imbalance penalties respectively; with the adopted $\alpha=1$ the weight reduces to $w^{\beta}\,|y-c|$.

The total objective is a weighted sum of the eight item-level losses,
\begin{equation}
\mathcal{L}_{\text{total}} = \sum_{q=1}^{8} \lambda_q \cdot \mathcal{L}_{\text{ImbOLL}}^{(q)}, \quad \lambda_q = \frac{|\tilde{\lambda}_q|}{\sum_{q'} |\tilde{\lambda}_{q'}|},
\end{equation}
where $\tilde{\lambda}_q$ are free parameters learned jointly with the model and normalised to sum to one at every step, so that the relative contribution of each item to the objective is learned rather than fixed in advance. They are initialised slightly higher for Q2 and Q6, the items most strongly associated with the total score.

\subsection*{Experimental setup}
\label{sec:experimental-design}

Each ablation row removes exactly one component from the full model. The rows remove participant-speech isolation (the original E-DAIC transcripts, interviewer turns included, with the same window repair), cross-modal attention (keys and values drawn from the query modality, so that depth and parameter count are unchanged and only the cross-modal information flow is removed; undefined for a single modality), the graded-response head (replaced by eight independent item heads on the same fused representation) and multi-task training (eight single-item models per seed, each trained on one item and summed at evaluation). To make each row attributable to a single component, all conditions are produced by one implementation on one shared feature cache, and share the same partition, optimiser, schedule, checkpoint selection rule and five random seeds; each row whose component is removed is initialised from unimodal trunks trained without that component. We report the test partition in the main text and the development partition in Supplementary Table~S8. The development partition is used for checkpoint selection and for the loss-hyperparameter search, so it is not an independent held-out estimate; it differs from the test partition in composition, having a lower label mean and variance and a mixture of Wizard-of-Oz and autonomous sessions rather than autonomous sessions alone, so an effect consistent in sign across the two is more likely to reflect the component than the particular 56 participants it was measured on.

\subsubsection*{Sequence handling}

Interviews differ in length, so each modality's utterance sequence is zero-padded to a common width of $T_{\max}=203$, the longest original transcript, and no interview is truncated. A padded batch can be given to a recurrent encoder in two ways. When the sequences are packed, the encoder is told the true length of each sequence and computes its recurrent state only over the real utterances. When they are not, it runs over the padded positions as well; the forward state is unaffected at the real utterances, but the backward state arriving at the first real utterance has already traversed every padded position, so it encodes how much padding followed and therefore the length of the interview, which correlates with severity on this corpus. Every model reported here packs its sequences, and pooled representations are averaged over the real utterances only. For the comparison reported under RQ2 we retrained the audio-only and video-only models with packing disabled and everything else held fixed, and additionally in a configuration that truncates sequences at 120 utterances and uses eight independent item heads.

\subsubsection*{Permutation test and attention diagnostics}

To measure how far a trained multimodal model depends on each input stream, we permuted the utterance sequences of one modality across the 56 test participants, so that each participant's prediction was computed from another participant's stream while the other streams, the trained weights and the decoding rule were left unchanged. 
Since expected-value totals are continuous and shift slightly under any input change, this test sums each item's most probable category (i.e., argmax) for easier interpretation. We report the percentage of test participants whose predicted total score differs from the unpermuted prediction, averaged over the five random seeds. At evaluation we also record the entropy of every self- and cross-attention module, expressed as a ratio to the entropy of a uniform distribution over the real utterances. In trained models every cross-attention module has a ratio of 1.000 on every seed, and no self-attention module falls below 0.55, since the BiLSTM outputs are small and nothing rescales the attention logits, so that each cross-modal module reduces to mean pooling over the utterances.

\subsubsection*{Implementation and model selection}

We implemented SIGMA in PyTorch. Unimodal models were trained first, and each multimodal model was initialised from the unimodal encoders trained with the same seed, with the text encoder kept frozen while the audio and video encoders, the fusion and the head were trained. We trained each model with five random seeds and report the mean of the five runs, which reduces the variance of single-seed estimates. Supplementary Table~S4 lists all architectural and optimisation settings. The code, feature caches, trained checkpoints and result tables are provided with the paper.

We selected the W-ImbOLL hyperparameters by grid search on the development set, evaluating $\alpha \in \{1.0,1.5,2.0\}$ and $\beta \in \{0.5,1.0\}$ with five random seeds per configuration (Supplementary Table~S5). The configuration $(\alpha=1.0,\beta=0.5)$ gave the highest mean development CCC ($0.592\pm0.030$ for the text-only model). Increasing $\alpha$ to 2.0 lowered CCC at $\beta=0.5$, from 0.592 to 0.557, and left it unchanged at $\beta=1.0$ (0.503 against 0.505), which suggests that a strong penalty on ordinal distance is unhelpful on a dataset of this size and imbalance. For each $\alpha$, the setting $\beta=0.5$ outperformed $\beta=1.0$, indicating that moderate class weighting is preferable. We therefore fixed $\alpha=1.0$ and $\beta=0.5$ for all subsequent experiments, consistent with prior work \cite{mandal2025enhancing}. Test data were not used for hyperparameter selection.

\subsubsection*{Evaluation metrics}

Following ref.~\citenum{mandal2025enhancing}, we evaluated performance on each partition using three regression metrics computed from the total PHQ-8 score. For $N$ samples with ground-truth scores $y_i$ and predictions $\hat{y}_i$, the root mean square error (RMSE) and mean absolute error (MAE) are
\begin{equation}
\text{RMSE} = \sqrt{\frac{1}{N} \sum_{i=1}^{N} (y_i - \hat{y}_i)^2}, \qquad \text{MAE} = \frac{1}{N} \sum_{i=1}^{N} |y_i - \hat{y}_i|.
\end{equation}
Both are non-negative, and lower values indicate better accuracy. We also report the concordance correlation coefficient (CCC), which measures agreement between predicted and ground-truth scores by combining correlation with differences in their means and variances:
\begin{equation}
\rho_c = \frac{2 \rho\, \sigma_y \sigma_{\hat{y}}}{\sigma_y^2 + \sigma_{\hat{y}}^2 + (\mu_y - \mu_{\hat{y}})^2},
\end{equation}
where $\rho$ is the Pearson correlation between $y$ and $\hat{y}$, $\mu_y$ and $\mu_{\hat{y}}$ are their means, and $\sigma_y$ and $\sigma_{\hat{y}}$ their standard deviations. CCC ranges from $-1$ to $1$, and higher values indicate stronger agreement \cite{lawrence1989concordance,mandal2025enhancing}. Each item score is decoded from the graded-response head as its expected value under the predicted category distribution, and the eight expected scores are summed to the total. We report expected-value decoding throughout, since it preserves the ordinal spacing of the categories.

\section*{Data availability}

E-DAIC can be requested from the USC Institute for Creative Technologies at \url{https://dcapswoz.ict.usc.edu/}, subject to the provider's access terms. This study generated no new data.

\section*{Code availability}

The SIGMA code, the utterance-level feature caches derived from E-DAIC, the trained checkpoints of every reported run and the scripts that regenerate the tables and figures are available at \url{https://github.com/wtgme/SIGMA}.

\section*{Ethics declarations}

We conducted a secondary analysis of the de-identified and publicly documented E-DAIC corpus \cite{gratch2014distress,devault2014simsensei,ringeval2019avec}. The USC Institute for Creative Technologies was responsible for the original data collection, informed consent and ethical approval, as reported in the corpus publications. This study collected no new human-participant data, involved no new recruitment or interaction with participants, and accessed no identifiable information.

\section*{Acknowledgements}

This study was funded by: (a) Maudsley Charity, (b) the National Institute for Health Research (NIHR) Biomedical Research Centre at South London and Maudsley NHS Foundation Trust and King's College London. T.W.\ is part-funded by the King's Together fund and MHaPS Early Career Researcher Award. R.D.\ is also funded by (a) Health Data Research UK; (b) The BigData@Heart Consortium, funded by the Innovative Medicines Initiative-2 Joint Undertaking under grant agreement No.\ 116074; and (c) The National Institute for Health Research University College London Hospitals Biomedical Research Centre. A.R.\ is funded by the UK Medical Research Council (grant DATAMIND, The Data Hub for Mental Health Informatics Research and Development, number MR/Z504816/1; and grant RE-STAR, Regulating Emotions: Strengthening Adolescent Resilience, number MR/W002493/1). The views expressed are those of the authors and not necessarily those of the NHS, the NIHR or the Department of Health. The funding bodies had no role in the study design; the collection, analysis or interpretation of data; the preparation of the manuscript; or the decision to submit the manuscript for publication.

\section*{Author contributions}

Z.T.\ designed and implemented the framework, conducted the experiments, analysed the findings and drafted the manuscript. T.W.\ conceived and supervised the study, contributed to its design and methodology, conducted the experiments and analysed the findings. K.Y.\ and L.Y.\ contributed to model design and analysis of the multimodal fusion experiments. Y.Z.\, R.Z.\, S.C.S.\ and J.D.\ contributed to clinical interpretation of the item-level PHQ-8 findings and to the study framing. Y.C.\ contributed to experimental design and validation. R.D.\ and A.R.\ provided methodological and informatics supervision. All authors critically revised the manuscript and approved the final version.

\section*{Competing interests}

The authors declare no competing interests.

\bibliography{sn-bibliography}

\end{document}